\documentclass[twocolumn,showpacs,amsmath,amstex,amssymb,mathfonts,prb]{revtex4-1}
\usepackage{amsthm,amsfonts,graphicx,verbatim}

\usepackage{dcolumn}   
\usepackage{verbatim}
\usepackage{tipa}
\usepackage{wasysym}
\usepackage{esint}
\usepackage{color}

\usepackage{amsmath}
\usepackage{amssymb}
\usepackage{amsthm}
\usepackage{amsfonts}
\usepackage{listings}
\usepackage{enumerate}
\usepackage{latexsym}
\usepackage{subfigure}

\begin{document}

\title{Analytic Wavefunctions for Neutral Bulk Excitations in Fractional Quantum Hall Fluids}

\author{ Bo Yang$^1$}
\affiliation{$^1$ Department of Physics, Princeton University, Princeton, NJ 08544, USA}

\date{\today}
\begin{abstract}
We show the model wavefunctions for the neutral collective modes in fractional quantum Hall (FQH) states have simple analytic forms obtained from judicially reducing the powers of selected pairs in the ground state Jastrow factor. This scheme of ``pair excitations" works for the magneto-roton modes of single-component Abelian and non-Abeliean FQH states, as well as the neutral fermion mode for the Moore-Read (MR) state. The analytic wavefunctions allow us to compute the ``quadrupole gap" of the magneto-roton mode in the thermodynamic limit, which was previously inaccessible to the numerics. The quadrupole gap is related to the fusion of the charges in the two-dimensional plasma picture, extending the plasma analogy to neutral excitations. A lattice diagrammatic method of representing these many-body wavefunctions and FQH elementary excitations is also presented.
\end{abstract}

\maketitle \vskip2pc
\section{Introduction}\label{sec_introduction}

The fractional quantum Hall effect (FQHE)\cite{tsg} is one of the prime examples where the strong interaction between electrons dictates the dynamics. A very fruitful approach for such a strongly correlated system is to look for model wavefunctions and model Hamiltonians that are adiabatically connected to experimentally accessible systems. For the ground states and charged excitations of FQHE at several filling factors, wavefunctions have compact analytic forms\cite{l,m,r}. We can thus infer the properties of FQHE in the thermodynamic limit by reinterpret the wavefunctions in analogy to a two-dimensional plasmas\cite{l}, or as conformal blocks of some special conformal field theories (CFT)\cite{m,bonderson}. In general for these approaches, incompressibility of FQHE is always assumed, and the dynamics of the bulk gapped excitations is not explicitly addressed.

Incompressibility of FQHE is defined by the neutral bulk collective excitations. Such excitations are important for understanding which topological phases of the FQHE can be stabilized, and are also very relevant to the recent development in fractional Chern insulator, where the issue of incompressibility and its mapping to the FQHE\cite{ss,sm,wjs,roy,lb,wrb,ltq} are areas of active research. The first formal treatment of the neutral excitations came from the single mode approximation (SMA) for the magneto-roton\cite{gmp} mode, where good model wavefunctions of density wave excitations can be constructed numerically up to the momentum of roton minimum, beyond which SMA is no longer valid\cite{yzz}. Following the experimental studies of the neutral excitations by several groups\cite{west1,foxon,cheng,west2,west3}, more recently the model wavefunctions are constructed numerically both for the magneto-roton modes in the Read-Rezayi (RR) series, and the neutral fermion mode\cite{moller} in the Moore-Read (MR) state, the latter reflecting its non-Abelian nature. One approach in constructing the model wavefunctions is to treat the neutral excitations as excitons of the composite fermions\cite{jain1,jain2,rodriguez}. Another approach is to apply the formalism of Jack polynomials\cite{bernevig} with appropriate root configurations and clustering properties\cite{bernevig1,yzz}. These wavefunctions are in principle good for the entire range of momenta. In practice, however, the long wavelength limit is not accessible due to the limitation by the system size. Interestingly, even though the underlying phenomenological pictures of the neutral excitations can be different, it is found numerically that both approaches mentioned above produce exactly the same set of model wavefunctions with very rich algebraic structures. Just like the ground states and bulk charged excitations of many FQH fluids, where model wavefunctions have natural forms with no variational parameters, in this paper we show how the model wavefunctions for the bulk neutral excitations can be formulated analytically, thereby unifying previous numerical works on this issue.

This paper is organized as follows: in Sec.~\ref{sec_wfconstruction} the model wavefunctions for both the magnetoroton mode and the neutral fermion mode are constructed explicitly, generalizing the Jastrow factor in the Laughlin wavefunction both at odd and even filling. In Sec.~\ref{sec_diagram} a diagrammatic scheme is introduced to represent the bulk neutral many-body wavefunctions in an intuitive way, which also shed light on the way these neutral excitations can be interpreted as ``elementary excitations" of the FQH fluid. In Sec.~\ref{sec_quadrupole} the compact analytic form of the model wavefunctions is exploited to calculate the thermodynamic energy gap of the neutral excitations in the long wavelength limit, which lies in the region inaccessible to numerical calculation. This also leads to an interesting connection of the FQHE bulk dynamics to the free energy cost of particle fusion in the plasma analogy. Sec.~\ref{sec_conclusion} gives the conclusion and discussion of the paper.

\section{Wavefunction Construction}\label{sec_wfconstruction}

Previous works on the numerical generation of the neutral excitation model wavefunctions are done mostly on spherical geometry, where only gapped bulk excitations are present for the incompressible phases. The bulk neutral excitations are also present on disk geometry, though the spectrum is more complicated due to the presence of edge excitations (more comments on it in Sec.~\ref{sec_conclusion}). In the thermodynamic limit the bulk excitations should be insensitive to the geometry of the Hall manifold. In fact even for finite systems, we can go to the ``conformal limit"\cite{bernevigconformallimit} on either sperical or disk geometry, by removing the single particle wavefunction normalization constant from the many-body wavefunctions; the resulting model wavefunctions are identical. For example, the Laughlin wavefunction of the two fermions at filling factor $\nu=1/3$ on the sphere is given by $\psi_s=\frac{1}{\sqrt{2}}\left(|1001\rangle-|0110\rangle\right)$, and on the disk it is given by $\psi_d=\frac{1}{2}|1001\rangle-\frac{\sqrt{3}}{2}|0110\rangle$. The string of binary numbers represents the orbital basis\cite{yzz}, where the $1's$ denote occupied orbitals and $0's$ denote unoccupied orbitals. On the sphere, the leftmost orbital is at the north pole, and the rightmost orbital is at the south pole; on the disk, the leftmost orbital is at the center of the disk, while the rightmost orbital is at the edge. For two electrons, the proper number of orbitals is \emph{four} to account for the shift on the sphere\cite{wen}. When the single particle normalization is removed, both wavefunctions lead to the familiar Laughlin wavefunction in the conformal limit $\psi_c=|1001\rangle-3|0110\rangle$, where the coefficient of the root configuration $|1001\rangle$ is normalized to unity. It is thus convenient to unambiguously rewrite the wavefunction with the un-normalized single particle wavefunction on the disk in the lowest Landau level (LLL), where the $n^{\text{th}}$ orbital is given by $z^{n-1}_i$(the Gaussian factor is omitted since it does not play any role here for FQHE), with holomorphic variables $z_i=\frac{1}{\sqrt{2}l_B}\left(x_i+iy_i\right)$. Here $x_i$ and $y_i$ are the coordinates in the Hall manifold, $l_B=\sqrt{\hbar/eB}$ is the magnetic length and $i$ is the particle index. Explicitly we have
\begin{eqnarray}\label{conversion}
|1001\rangle\sim z_1^3-z_2^3,\quad|0110\rangle\sim z_1z_2^2-z_1^2z_2
\end{eqnarray}
and $\psi_c\s=\left(z_1-z_2\right)^3$.

Having identified the relationship between many-body wavefunctions on different geometries, in this paper all analytic wavefunctions are presented as polynomials of $z_i$. Thus in principle these model wavefunctions are for neutral excitations on the disk, but they can be easily converted into model wavefunctions on the sphere by multiplying appropriate spherical single particle normalization factors. In particular, after the conversion the model wavefunctions in this paper are \emph{identical} to those generated numerically in \cite{jain1,jain2,rodriguez,yzz}. On the disk each model wavefunction is labeled by its total angular momentum about the z-axis perpendicular to the disk, i.e.$\delta L_z$ measured from the ground state (with the ground state having $\delta L_z=0$). The neutral excitations are states with negative $\delta L_z=-N$(states with positive $\delta L_z$ contain gapless edge excitations). When the highest weight condition is imposed\cite{hweight}, these states correspond to the highest weight state on the sphere in the total angular momentum sector $L=N$, with all the quasiparticles piled at the north pole\cite{yzz}. Since the mapping from the disk wavefunctions to the sphere wavefunctions is unambiguous, in this paper all model wavefunctions are labeled by $L$ instead of $\delta L_z$, to facilitate comparison with numerical works on the spherical geometry in the literature.

On the sphere the ground state is the Laughlin wavefunction in total angular momentum $L=0$ sector. The corresponding model Hamiltonian on the disk made of Haldane pseudopotentials\cite{haldane} is given by $V=\sum_{i<j}V_{ij}$, with
\begin{eqnarray}\label{vij}
V_{ij}=\int \frac{d^2ql_B^2}{2\pi}\sum_{n=0}^{m-1}L_n(q^2l_B^2)e^{-\frac{1}{2}q^2l_B^2}e^{i\vec q\cdot (\vec R_i-\vec R_j)}
\end{eqnarray}
where $L_n(x)$ is the $n^{\text{th}}$ Laguerre polynomial and $\vec R_i$ is the guiding center coordinate of the $i^{\text{th}}$ particle. Physically, $V_{ij}$ is the short range interaction that projects into the two-body Hilbert space with the relative angular momentum smaller than $m$. 

We now present the wavefunctions of the neutral excitations for the fermionic Laughlin state at filling factor $\nu=1/m$ in the lowest Landau level (LLL), where $m$ is odd. The family of the neutral excitations at different angular momentum sectors is as follows:
\footnotesize
\begin{widetext}
\begin{eqnarray}\label{collect}
&&\mathcal A[(z_1-z_2)^{m-2}\prod'_{i<j}(z_i-z_j)^m]\qquad L=2\nonumber\\
&&\mathcal A[(z_1-z_2)^{m-2}(z_1-z_3)^{m-1}\prod'_{i<j}(z_i-z_j)^m]\qquad L=3\nonumber\\
&&\mathcal A[(z_1-z_2)^{m-2}(z_1-z_3)^{m-1}(z_1-z_4)^{m-1}\prod'_{i<j}(z_i-z_j)^m]\qquad L=4\nonumber\\
&&\qquad\qquad\qquad\qquad\qquad\qquad\qquad\vdots
\end{eqnarray}
\end{widetext}
\normalsize
Here $\mathcal A$ indicates antisymmetrization over all particle indices, and the prime sign on $\prod'_{i<j}$ indicates the product of only pairs $\{ij\}$ that do not appear in the prefactors to the left of it. For example, in Eq.(\ref{collect}) the product in the $L=2$ wavefunction does not contain $(z_1-z_2)^m$.

An explanation of Eq.(\ref{collect}) is in order here. The $L=2$ state, which is the quadrupole excitation in the thermodynamic limit\cite{yzz}, is obtained from the ground state by reducing the power of one pair of particles (which we can choose arbitrarily as particle $1$ and $2$ because of antisymmetrization) by \emph {two}, followed by antisymmetrizing over all particles. This scheme naturally forbids an $L=1$ state by the pair excitation, since if we reduce the power of one pair of particles by \emph {one}, antisymmetrization kills the state. The $L=3$ state is generated by pairing particle $1$ with another particle (which we can arbitrarily choose as particle $3$) and reducing their pair power by one. It is now clear how the modes in other momentum sectors $L=4,5,\cdots$ are generated. Naturally for a total of $N_e$ particles, the family of neutral excitation modes ends at $L=N_e$, agreeing with the numerical schemes in the literature. We have numerically checked for different system sizes that all wavefunctions in Eq.(\ref{collect}) agrees exactly with those generated in\cite{jain1, jain2, rodriguez, yzz}. Indeed all wavefunctions here satisfy the highest weight condition, and the states relax to the ground state far away from the excited pairs; these are exactly the conditions we used to numerically generate the unique model wavefunction in each momentum sector\cite{yzz}.

The same scheme applies to the MR state. It is instructive to first see how the MR ground state is obtained. The Laughlin wavefunction at half filling is given by the Jack polynomial\cite{bernevig} $J^{-2}_{1010101\cdots}(z_i)=\prod_{i<j}(z_i-z_j)^2$. For fermions this is not a valid state; instead the ground state was constructed by a pairing mechanism\cite{m}, which is also a Jack polynomial $J^{-3}_{1100110011\cdots}$. Explicitly in the wavefunction, the pairing reduces the power of each pair of particles by \emph{one}. For $2n$ particles, the antisymmetrization reproduces the Pfaffian up to a constant as follows:

\footnotesize
\begin{eqnarray}\label{pairing}
&&\text{Pf}\left(\frac{1}{z_i-z_j}\right)\prod_{i<j}(z_i-z_j)^2\nonumber\\
&&\sim\mathcal A[(z_1-z_2)(z_3-z_4)\cdots(z_{2n-1}-z_{2n})\prod'_{i<j}(z_i-z_j)^2]
\end{eqnarray}
\normalsize

Again the prime sign in the second line indicates products of only pairs $\{ij\}$ other than $\{1,2\},\{3,4\},\cdots \{2n-1,2n\}$ appearing in the prefactor.  The explicit use of antisymmetrization instead of the Pfaffian allows us to naturally extend to the case with an odd number of particles: starting from the Bosonic Laughlin wavefunction at half filling, every two particles form a pair except for just one particle. Naturally the ``ground state" of the neutral fermion mode is given by
\small
\begin{eqnarray}\label{opairing}
\mathcal A[(z_1-z_2)(z_3-z_4)\cdots(z_{2n-1}-z_{2n})\prod'_{i<j}(z_i-z_j)^2]
\end{eqnarray}
\normalsize

Note both $i,j$ in $\prod'_{i<j}$ runs from $1$ up to $2n+1$, with pairs appearing before  $\prod'_{i<j}$ excluded. Though we can no longer represent Eq.(\ref{opairing}) as a Pfaffian, comparing the antisymmetrized products we can see Eq.(\ref{opairing}) is really the same as that of Eq.(\ref{pairing}), only with an odd number of particles. For the model three-body Hamiltonian, this is a zero-energy abelian quasihole state $J^{-3}_{1100110011\cdots 0011001}$ in the angular momentum sector $L=\frac{1}{2}(N_e-1)$. The magneto-roton mode and the neutral fermion mode are obtained from Eq.(\ref{pairing}) and Eq.(\ref{opairing}) respectively by reducing the powers in the Jastrow factor the same way as what is done for the Laughlin state.

To write down all the analytic wavefunctions shown above in a more formal way, we define $\mathcal P_{ij}=\frac{1}{z_i-z_j}$. Notice the Pfaffian for $2n$ particles can be written as $\text{Pf}\left(\frac{1}{z_i-z_j}\right)\sim\mathcal A[\mathcal P^{(2n)}]$, where $\mathcal P^{(2n)}=\mathcal P_{12}\mathcal P_{34}\cdots\mathcal P_{2n-1,2n}$. The magneto-roton mode for the Laughlin state in the $L=k+2$ sector is given by
\begin{eqnarray}\label{laughlinm}
\psi_l^{L=k+2}=\prod_{i<j}^{N_e}(z_i-z_j)^m\mathcal S[\mathcal P_{12}^2\mathcal P_{13}\cdots\mathcal P_{1,2+k}]
\end{eqnarray}
where $\mathcal S$ is the symmetrization over all the particle indices. From the Bosonic Laughlin wavefunction at filling factor $1/2$ we can impose pairing to obtain
\begin{eqnarray}\label{mrg}
\psi_{\text{mr}}=\prod_{i<j}^{N_e}(z_i-z_j)^2\mathcal A[\mathcal P^{(2n)}]
\end{eqnarray}
For an even number of electrons we have $N_e=2n$ and Eq.(\ref{mrg}) is the MR ground state. The magneto-roton mode for the MR state in the $L=k+2$ sector is given by
\small
\begin{eqnarray}\label{mrm}
\psi_{\text{mr}}^{L=k+2}=\prod_{i<j}^{N_e}(z_i-z_j)^2\mathcal A[\mathcal P^{(2n)}\mathcal P_{13}^2\mathcal P_{15}\cdots \mathcal P_{1,3+2k}]
\end{eqnarray}
\normalsize
 For an odd number of electrons we have $N_e=2n+1$ and Eq.(\ref{mrg}) is the MR quasihole state of Eq.(\ref{opairing}). The neutral fermion mode in the $L=3/2+k$ sector is given by
\footnotesize
\begin{eqnarray}\label{mrn}
\psi_{\text{mr}}^{L=\frac{3}{2}+k}=\prod_{i<j}^{N_e}(z_i-z_j)^2\mathcal A[\mathcal P^{(2n)}\mathcal P_{N_e,1}^2\mathcal P_{N_e,3}\cdots\mathcal P_{N_e,1+2k}]
\end{eqnarray}
\normalsize

All the model wavefunctions shown so far satisfy the highest weight condition. On the disk these states describe neutral systems with no center of mass rotation, of which the ground states are just special cases with no symmetrized/antisymmetrized singular factors multiplying to the Jastrow factor.

\section{Diagrammatic Representation}\label{sec_diagram}

\begin{figure}[h!]
\includegraphics[width=3.5cm,height=1.5cm]{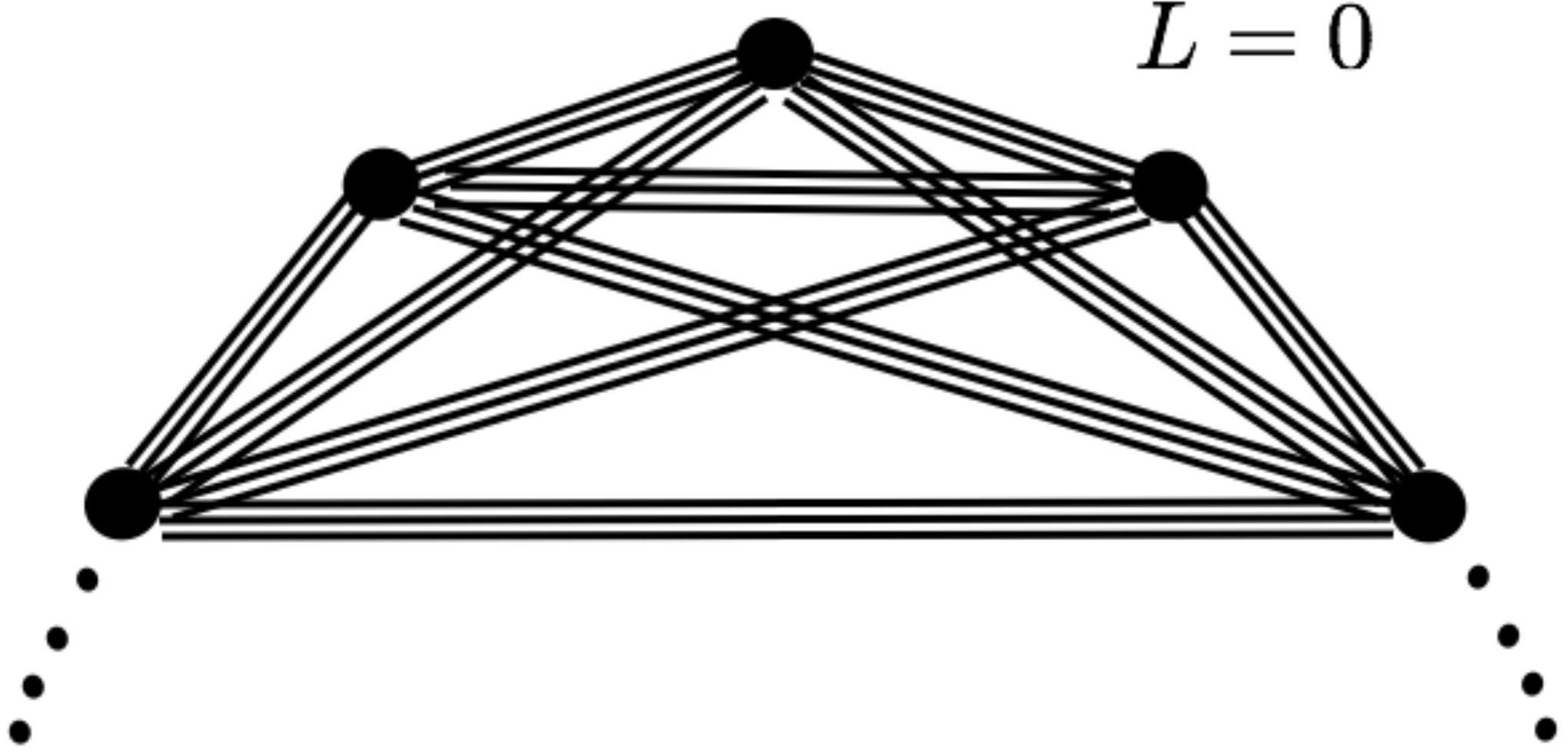}
\caption{For $N_e$ particles, the lattice can be viewed as an N-gon, with three bonds connecting every pair of vertices.}
\label{fig:laughlin0}
\end{figure}

An intuitive way to visualize the family of neutral excitations is to map the particles onto a lattice, where each lattice site represents a particle. Since for FQHE we have a quantum fluid instead of a solid, every two lattice sites interact with each other. The number of bonds between each pair of lattice sites equal to the power of the pair of particles in the wavefunction. As an example we consider the simpliest Laughlin state at $\nu=1/3$, so for the ground state every two lattice sites are connected by three bonds, as shown in Fig.~\ref{fig:laughlin0}.

\begin{figure}[h!]
\includegraphics[width=5.5cm,height=3.5cm]{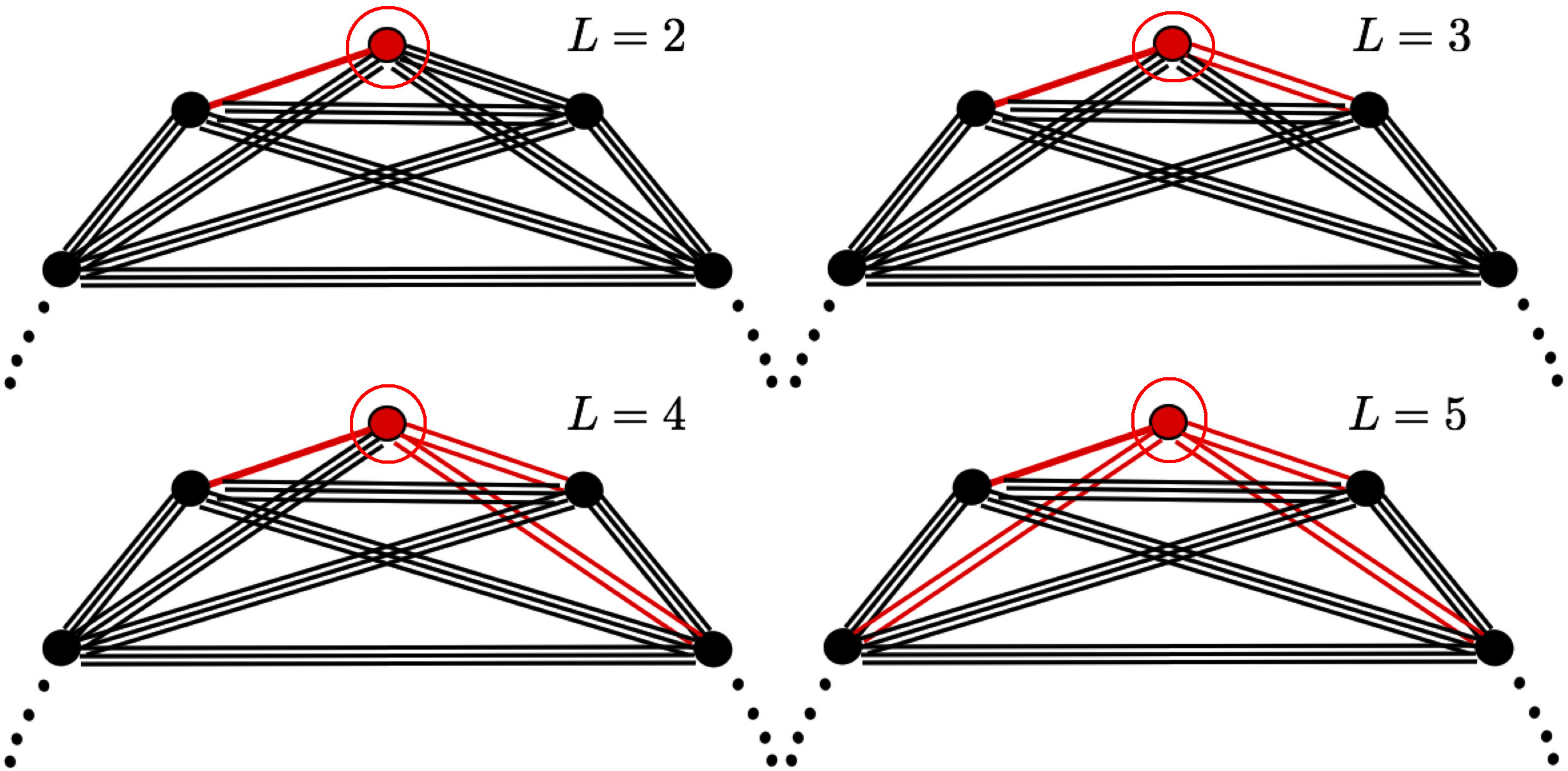}
\caption{(Color Online) Collective modes from $L=2$ to $L=5$, where the change of bonds are highlighted with red color and the involved lattice cite is circled.}
\label{fig:laughlin25}
\end{figure}

The neutral excitations are obtained by breaking the bonds between lattice sites, as shown in Fig.~\ref{fig:laughlin25}. We can view the entire family of the neutral excitations as the elementary excitations centered around a single red lattice site. Note the lattice pattern uniquely defines the many-body wavefunction, and different types of ``elementary excitations" can be identified with different patterns of bond-breaking around a single lattice site with the red color (and also circled). 

\begin{figure}[h!]
\includegraphics[width=5.5cm,height=1.5cm]{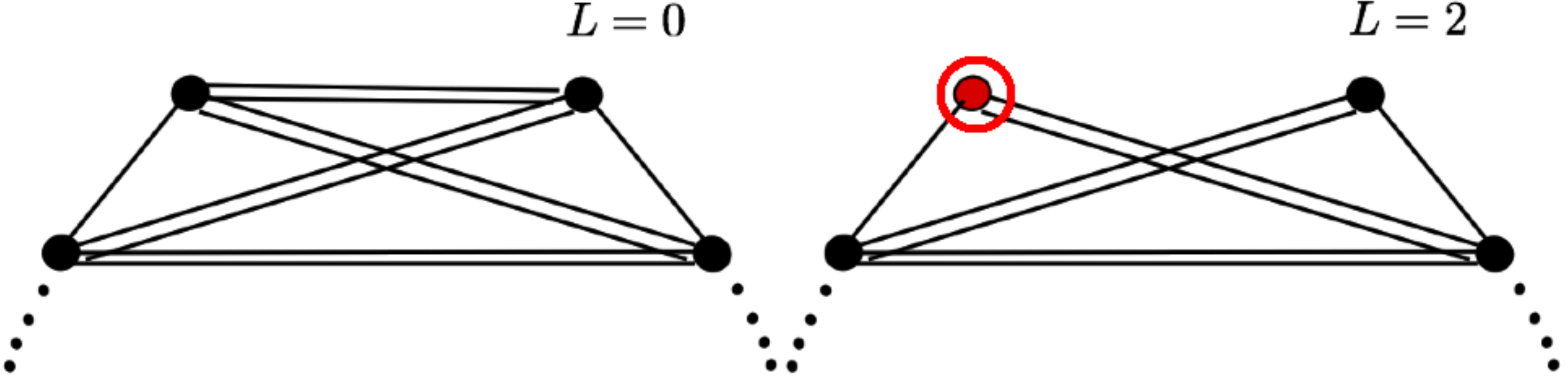}
\caption{(Color Online) The lattice configuration of the ground state $L=0$ and the first collective mode $L=2$. Consecutive collective modes can be obtained by breaking one of the \emph {double bonds} connecting the red (circled) lattice site to some other site.}
\label{fig:mr02}
\end{figure}

\begin{figure}[h!]
\includegraphics[width=3.5cm,height=4.5cm]{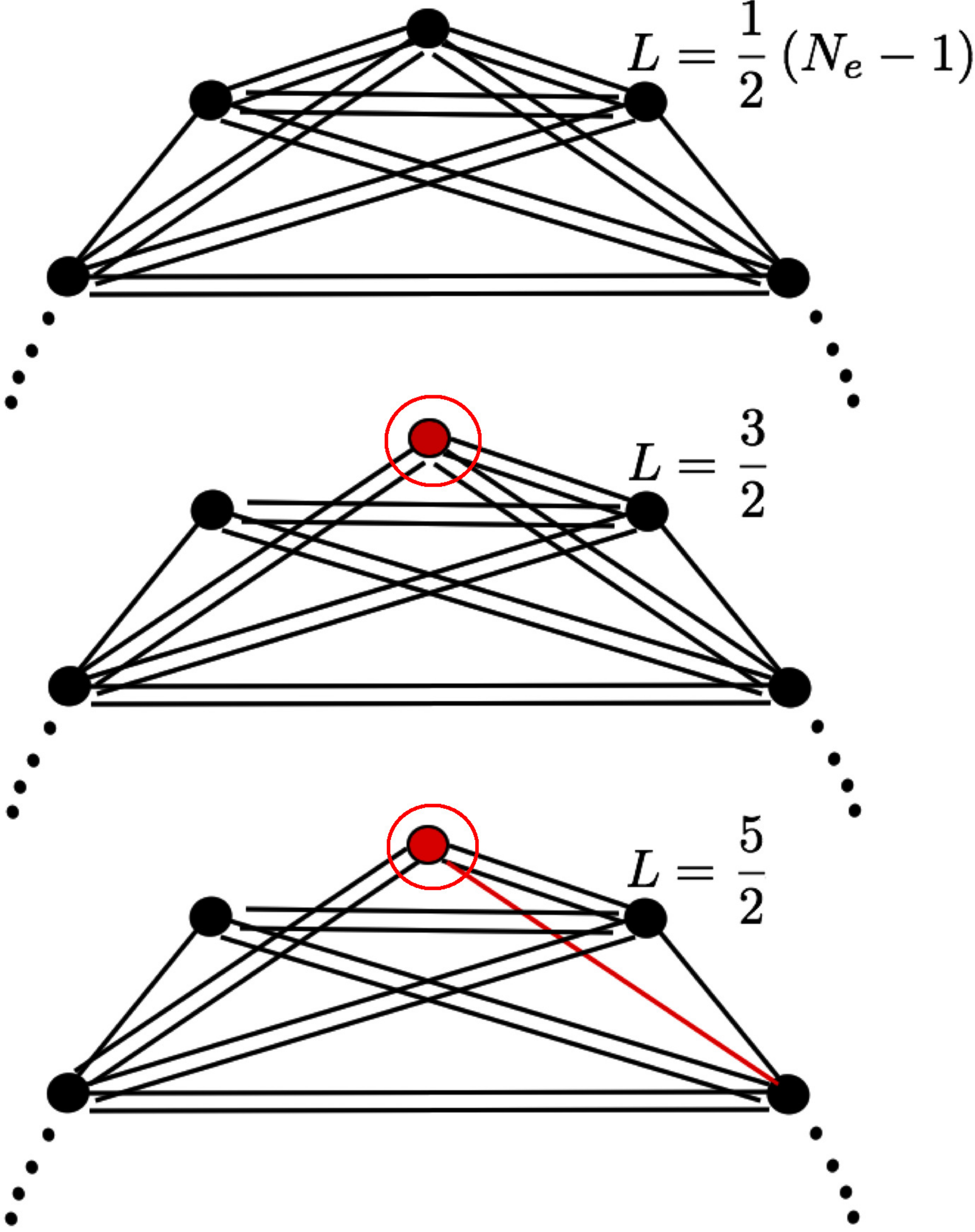}
\caption{(Color Online) The lattice configuration of the zero mode quasihole state $L=\frac{1}{2}\left(N_e-1\right)$ and the first two neutral fermion modes at $L=\frac{3}{2}\text{ and }L=\frac{5}{2}$. Consecutive collective modes can be obtained by breaking one of the \emph {double bonds} connecting the red lattice site to some other site.}
\label{fig:mrn02}
\end{figure}

The neutral excitations in the MR states can be represented similarly. For the MR ground state with even number of particles, every paired particles corresponds to a pair of lattice sites with only one bond connecting them. This can be seen in the first line of Eq.(\ref{pairing}), as the paired coordinates only have power 1 in the Jastrow factor. All other pairs of lattice sites have two bonds connecting them (see the left diagram of Fig.(\ref{fig:mr02}), where only four sites are explicitly shown). For an odd number of particles, the zero-energy quasihole state is obtained from the MR ground state lattice by adding one more lattice site that connects every other lattice site with two bonds (See the first line of Eq.(\ref{opairing}), and the top diagram of Fig.(\ref{fig:mrn02})).

The lattice representations of the magneto-roton modes for the MR state are given in Fig.(\ref{fig:mr02}) and those of the neutral fermion mode are given in Fig.(\ref{fig:mrn02}).

The lattice diagrams give a one-to-one mapping to the many-body wavefunctions; one would also conjecture the diagrams are useful in determining the many-body wavefunctions of multi-roton excitations, with the same bond-breaking pattern around more than one lattice sites. One should note, however, the different many-body wavefunctions corresponding to different lattice diagrams \emph{may not} be linearly independent, and this is a subject requiring further research. All the wavefunctions presented in this paper are linearly independent because they are in different angular momentum sectors.

\section{Energy Gap of Quadrupole Excitations}\label{sec_quadrupole}

For systems with finite number of particles, the analytic wavefunctions presented in the previous two sections do not seem to be advantageous in calculating either the density profile or the variational energy, unless there is an ingenious way of implementing Monte Carlo techniques based on the explicit analytic form. In the thermodynamic limit where the number of particles $N_e\rightarrow\infty$, the spherical and disk geometry are equivalent, at least as far as the bulk neutral excitations are concerned. One can convert the total angular momentum $L$ of a state on the sphere to the linear momentum $k$ on the disk by $k=L/\sqrt{S}$, where $S$ is the monopole strength at the center of sphere\cite{haldane}. In the thermodynamic limit $S\rightarrow\infty$. Thus any state of finite $L$ on the sphere corresponds to a state with linear momentum $k\rightarrow 0$ on the plane, when we take the limit $N_e\rightarrow\infty$. In this way, the analytic wavefunctions are useful in calculating the thermodynamic neutral excitation gap in the long wavelength limit, or the so called ``quadrupole gap". Writing $\psi_l^{L=N}=\langle z_1,\cdots z_{N_e}|\psi_l^{L=N}\rangle$, for the Laughlin state, the energy gap is given by
\begin{eqnarray}\label{energy}
\epsilon_{k\rightarrow 0}=\lim_{N_e\rightarrow\infty}\frac{\langle\psi^{L=N}_l|V|\psi^{L=N}_l\rangle}{\langle\psi^{L=N}_l|\psi^{L=N}_l\rangle}
\end{eqnarray}

Physical arguments above lead to the conjecture that Eq.(\ref{energy}) is independent of $N$, and it would be nice to see a rigorous mathematical proof. We already know \cite{yzz} that in the $L=2$ and $L=3$ sector, the single mode approximation (SMA), which generates model wavefunctions via density wave excitations, is exact for the magneto-roton model wavefunctions on the sphere. Thus in the thermodynamic limit, the analytic wavefunctions on the disk labeled by $L=2,3$ are also identical to those generated by SMA in the limit $k\rightarrow 0$. It is thus easiest to evaluate Eq.(\ref{energy}) in the sector $L=2$. Defining the guiding center ladder operators as $b^\dagger_i=z_i,b_i=\partial_{z_i}$, and denoting the Laughlin wavefunction as $\psi_l=\prod_{i<j}(z_i-z_j)^m$, we have
\begin{eqnarray}\label{sma}
\psi^{L=2}_l&=&\frac{1}{2m(m-1)}\sum_i(b_i)^2\psi_l
\end{eqnarray}
In the thermodynamic limit, the normalization constant of the $L=2$ mode (and also for $L=3$ mode) is thus related to the long wavelength expansion of the ground state guiding center structure factor:
\begin{eqnarray}\label{sq}
S_q&=&\frac{1}{N_\phi}\left(\langle\delta\bar\rho_q\delta\bar\rho_{-q}\rangle_0-\langle\delta\bar\rho_q\rangle_0\langle\delta\bar\rho_{-q}\rangle_0\right)\nonumber\\
&=&-\frac{\bar s}{4m}(g^{ab}q_aq_b)^2+O(q^6)
\end{eqnarray}
where $\bar s=\frac{1-m}{2}$ is the guiding center spin\cite{gmp,haldane}, and $g^{ab}$ is the guiding center metric\cite{yzrh}. We thus have 
\begin{eqnarray}\label{denominator}
\langle\psi^{L=2}_l|\psi^{L=2}_l\rangle=-\frac{\bar sN_e}{2m^2(m-1)^2}
\end{eqnarray}
The numerator of Eq.(\ref{energy}) can be calculated with the plasma analogy. Note in Eq.(\ref{collect}), each wavefunction only has one pair of particles with relative angular momentum smaller than $m$. After antisymmetrization all particle indices are equivalent, so we only need to look at the action of $V_{12}$ on $\bar\psi_l^{\{12\}}$ with
\footnotesize
\begin{eqnarray}\label{psi12}
&&\langle z_1,\cdots, z_{N_e}|\bar\psi_l^{\{12\}}\rangle\nonumber\\
&&=(z_1-z_2)^{m-2}\prod_{2<i}^{N_e}(z_1-z_i)^m(z_2-z_i)^m\prod_{2<i<j}^{N_e}(z_i-z_j)^m
\end{eqnarray}
\normalsize
which is equivalent to the product within the antisymmetrization of the $L=2$ wavefunction  in Eq.(\ref{collect}). Writing $\bar z_{12}=\frac{1}{\sqrt{2}}\left(z_1+z_2\right), z_{12}=\frac{1}{\sqrt{2}}\left(z_1-z_2\right)$, with some simple algebra we obtain
\begin{eqnarray}\label{vpsi}
&&\langle z_1,\cdots, z_{N_e}|V_{12}|\bar\psi_l^{\{12\}}\rangle\nonumber\\
&&=z_{12}\prod_{2<i}^{N_e}\left(\frac{1}{\sqrt{2}}\bar z_{12}-z_i\right)^{2m}\prod_{2<i<j}^{N_e}(z_i-z_j)^m
\end{eqnarray}

Treating $z_{12}$ and $\bar z_{12}$ as independent particle variables, we can integrate out $z_{12}$ and the numerator of Eq.(\ref{energy}) is given by
\small
\begin{eqnarray}\label{onepair}
\langle\psi^{L=2}_l|V|\psi^{L=2}_l\rangle&=&\frac{N_e(N_e-1)}{2}\cdot\frac{\bar{\mathcal{N}}^2}{\mathcal N^2}
\end{eqnarray}
\normalsize
where $\mathcal N$ is the normalization constant of the Laughlin state $\psi_l$, and $\bar{\mathcal N}$ is the norm of the following wavefunction:
\small
\begin{eqnarray}\label{numerator}
\bar\psi=\prod_{i=2}^{N_e-1}\left(\frac{1}{\sqrt{2}}z_1-z_i\right)^{2m}\prod_{1<i<j<N_e-1}(z_i-z_j)^m
\end{eqnarray}
\normalsize
obtained from Eq.(\ref{vpsi}), which can be evaluated as the free energy of two-dimensional one-component plasma (OCP) on a disk with radius $R^2=\frac{mN_e}{2}$ and elementary charge $e=2\sqrt{\pi mk_BT}$, where particle $1$ interacts with the rest of the particles with charge $2e$. We thus obtain
\begin{eqnarray}\label{energyr}
\epsilon_{k\rightarrow 0}=-\frac{2^mm(m-1)^2}{\pi\bar s}e^{-\frac{\mathcal F_2-\mathcal F}{k_BT}}
\end{eqnarray}

Both $\mathcal F_2\text{ and }\mathcal F$ are the free energies of OCP in the thermodynamic limit ($N_e\rightarrow\infty$), where $\mathcal F$ is for $N_e$ particles, each with charge $e$ with logarithmic two-body interactions together with a neutralizing background of radius $R$; for $\mathcal F_2$, we have the same neutralizing background, but with $N_e-2$ particles of charge $e$, and exactly one particle with charge $2e$. Thus $\mathcal F_2-\mathcal F$ is the free energy cost of fusing two particles of charge $e$ to create a particle of charge $2e$, which is an $O(1)$ effect. The denominator is proportional to $\bar s$, which is negative by convention and it determines the overall strength of the quadrupole gap. Note for integer QH $\bar s=0$, and the quadrupole gap goes to infinity. This should be the case since the guiding center degrees of freedom are frozen out.

Similar calculations can be carried out for the magneto-roton mode in the MR state. Analogous to Eq.(\ref{sma}) we have $\psi^{L=2}_{\text{mr}}=\frac{1}{24}\sum_ib_i^2\psi_{mr}$,  and in the long wavelength limit we have
\begin{eqnarray}\label{energymr}
\epsilon^{\text{mr}}_{k\rightarrow 0}=-\frac{24}{\pi\bar s_{\text{mr}}}e^{-\frac{\mathcal F_3-\mathcal F_{\text{II}}}{k_BT}}
\end{eqnarray}
where  $\bar s_{\text{mr}}=-2$ is the guiding center spin for the MR state, and $\mathcal F_{\text{II}}$ is the standard two-component plasma free energy for the MR ground state\cite{bonderson}. The charge for the attractive interaction between the two components is given by $Q_1=\pm \sqrt{3k_BT}$, while the charge for the interaction between one component and the neutralizing background is given by $Q_2=2\sqrt{k_BT}$. $\mathcal F_3-\mathcal F_{\text{II}}$ is the free energy cost of fusing \emph {three particles} for each component to create one particle with charge $3Q_2$ but with the same $\pm Q_1$. 

The evaluation of the long wavelength gap of the neutral fermion mode is less transparent. The difficulty lies with evaluating the normalization constant of $\psi^{L=\frac{3}{2}}_{\text mr}$. There is no known SMA analogy for the neutral fermion mode, and it is not known if in the thermodynamic limit the gap should be inversely proportional to the guiding center spin. On the other hand $\langle\psi^{L=\frac{3}{2}}_{\text{mr}}|V_{\text{3bdy}}|\psi^{L=\frac{3}{2}}_{\text{mr}}\rangle$ can be mapped to 2-component plasma as well, and we obtain
\begin{eqnarray}\label{energymrnf}
\bar\epsilon^{mr}_{k\rightarrow 0}\sim e^{-\frac{\bar{\mathcal F}_3-\bar{\mathcal F}_{\text{II}}}{k_BT}}
\end{eqnarray}

Here $\bar{\mathcal F}_{\text{II}}$ is the free energy of the 2-component plasma similar to that of $\mathcal F_{\text{II}}$ with only one difference: there is exactly one \emph {more} particle carrying charge $Q_2$ that interacts with the neutralizing background, and its $Q_1$ charge is zero. This is how an unpaired fermion in the MR state is interpreted in the plasma analogy. Furthermore, $\bar{\mathcal F}_3-\bar{\mathcal F}_{\text{II}}$ is the energy cost of fusing the unpaired fermion with one pair of two other fermions, creating a particle with charge $Q_2=6\sqrt{k_BT}$ but again with \emph zero $Q_1$. The calculation of the prefactor in Eq.(\ref{energymrnf}) is not yet known.

\section{Conclusion and Discussion}\label{sec_conclusion}

In conclusion, analytic wavefunctions for both the magneto-roton modes and the neutral fermion modes are presented. The energy gap of the quadrupole excitation in the thermodynamic limit can be related to the free energy cost of the fusion of charges in the plasma energy, and is inversely proportional to the guiding center spin which characterizes its topological order. This is the first time that the plasma analogy is extended to neutral excitations of FQHE, and the analogy not only applies to the wavefunctions, but also to the dynamics as well. Since the neutral excitations in the long wavelength limit is buried in the multi-roton continuum, it is important to calculate the decay rate of these neutral modes. Numerical calculation has been performed to show that even in the continuum the decay rate of the collective mode is very small. This opens up the possibility of experimental detection of these modes. A more detailed analysis of the decay rate of collective neutral modes will be presented elsewhere\cite{yz}.

The neutral excitations in the single component FQHE can now be understood in several coherent framework, at least for the Laughlin and Moore-Read states, with possible generalization to the entire Read-Rezayi series. The composite fermion picture maps FQHE to the IQHE of the particle-vortex composite, and in this framework the neutral excitations are excitons of composite fermions. The Jack polynomial formalism enables us to describe the wavefunctions of the ground states, the quasihole and quasiparticle states, as well as the neutral excitations in a unified way with root configurations and squeezed basis constrained by the clustering properties. It is now satisfactory to see that compact analytic real space wavefunctions in electron coordinates, which initiated the theoretical understandings of FQHE, can now be extended from ground states and charged bulk excitations to include neutral bulk excitations. Some questions still remain on if the neutral excitations proposed so far completely describes the energy spectrum of FQHE. Experimental measurements on the Laughlin state\cite{west3} suggests a splitting of the collective modes in the long wavelength limit, with theoretical explanations proposed from a hydrodynamic point of view\cite{vignale}, and the composite fermion point of view\cite{jain2mode}. It would be interesting to see if the lattice diagram introduced in Sec.~\ref{sec_diagram} can be generalized to produce suitable analytic wavefunctions that describes multi-roton excitations and the splitting of the collective modes as well.

It is well-known in the literature that the wavefunctions of the gapless edge excitations on the disk can be obtained by multiplying the ground state with symmetric polynomials. With model Hamiltonians these are the zero energy states in the positive $\delta L_z$ angular momentum sectors\cite{wen,stone}. For the Moore-Read state, in addition to the charge sector generated by the symmetric polynomials, there are also edge excitations obtained from the statistical sectors via inserting Majorana fermions\cite{readedge}. The analytic wavefunctions of these states are known explicitly. One can also generate wavefunctions by similar operations not only on the ground state, but also on the bulk neutral excitations obtained in this paper. These wavefunctions describe states such that each contains both bulk and edge excitations. We call these roton-edge excitations, which explain the gapped low-lying multitude of states below the multi-roton gap in disk geometry. Recent studies show\cite{boyang} that for the Laughlin state, each bulk neutral excitation generates a branch of quasi-degenerate roton-edge excitations with the same Virasoro counting as the zero-energy edge states. For the Moore-Read state, however, the counting of the roton-edge states seem different because of the lack of the linear independence between states in the same momentum sector, possibly due to the non-abelian nature of the FQH fluid.

{\sl Acknowledgements}. I would like to thank F.D.M Haldane for many useful discussions, and Zlatko Papic for help in doing numerical evaluations. I also thank Chris Laumann for useful discussions and pointing me to the work of \cite{bonderson}. This work was supported by DOE grant DE-SC$0002140$ and NSS Scholarship by ASTAR.


\begin{thebibliography}{99}


\bibitem{tsg}
D. C. Tsui, H. L. Stormer, and A. C. Gossard, Phys. Rev. Lett. {\bf 48}, 1559 (1982).

\bibitem{gmp}
S. M. Girvin, A. H. MacDonald and P. M. Platzman, Phys. Rev. Lett. {\bf 54}, 581 (1985); Phys. Rev. B {\bf 33}, 2481 (1986).

\bibitem{ss}
S.A. Parameswaran, R. Roy and S.L. Sondhi, Phys. Rev. B. {\bf 85}, 241308 (2012).

\bibitem{sm}
T. Scaffidi and G. Moller, Phys. Rev. Lett. {\bf 109} 246805 (2012).

\bibitem{roy}
Rahul Roy, arXiv. 1208.2055

\bibitem{lb}
Zhao Liu and E.J. Bergholtz, Phys. Rev. B {\bf 87}, 035306 (2013).

\bibitem{wjs}
Ying-Hai Wu, J.K. Jain and Kai Sun, Phys. Rev. B. {\bf 86}, 165129 (2012).

\bibitem{wrb}
Yang-Le Wu, N. Regnault and B. A. Bernevig, Phys. Rev. Lett. {\bf 110}, 106802 (2013).

\bibitem{ltq}
Ching Hua Lee, R. Thomale and Xiao-Liang Qi, arXiv: 1207.5587

\bibitem{west1}
A. Pinczuk, B.S. Dennis, L.N. Pfeiffer and K.W. West, Phys. Rev. Lett. {\bf 70}, 3983 (1993).

\bibitem{foxon}
C.J. Mellor et al, Phys. Rev. Lett. {\bf 74}, 2339 (1995).

\bibitem{cheng}
U. Zeitler et al, Phys. Rev. Lett. {\bf 82}, 5333 (1999).

\bibitem{west2}
Moonsoo Kang, A. Pinczuk, B.S. Dennis, L.N. Pfeiffer and K.W. West, Phys. Rev. Lett. {\bf 86}, 2637 (2001).

\bibitem{west3}
C. F. Hirjibehedin, Irene Dujovne, A. Pinczuk, B. S. Dennis, L. N. Pfeiffer and K. W. West, Phys. Rev. Lett. {\bf 95}, 066803 (2005).

\bibitem{l}
R.B. Laughlin, Phys. Rev. Lett. {\bf 50}, 1395 (1983).

\bibitem{m}
G. Moore and N. Read, Nucl. Phys. B {\bf 360}, 362 (1991).

\bibitem{r}
N. Read and E. Rezayi, Phys. Rev. B {\bf 59}, 8084 (1999).

\bibitem{moller}
G. Moller, A. Wojs and N.R. Cooper, Phys. Rev. Lett. {\bf 107}, 036803 (2011)

\bibitem{bernevig1}
B. A. Bernevig and F. D. M Haldane, Phys. Rev. Lett. {\bf 102}, 066802 (2009)

\bibitem{bernevig}
B. A. Bernevig and F. D. M Haldane, Phys. Rev. Lett. {\bf 100}, 246802 (2008)

\bibitem{bernevigconformallimit}
R. Thomale, A. Sterdyniak, N. Regnault and B.A. Bernevig, Phys. Rev. Lett. 104, 180502 (2010).

\bibitem{hweight}
The center of mass angular momentum raising and lowering operators are given by $L^+=\sum_iz_i$ and $L^-=\sum_i\partial_{z_i}$, which both commutes with the Hamiltonian due to rotational invariance. The highest weight condition of a state $\psi$ is defined as $L^-\psi=0$, so the state $\psi$ does not have center of mass rotation.

\bibitem{jain1}
R.K. Kamilla, X.G. Wu and J.K. Jain, Phys. Rev. B. {\bf 54}, 4873 (1996)

\bibitem{jain2}
G.J. Sreejith, A. Wojs and J.K. Jain, Phys. Rev. Lett. {\bf 107} 136802 (2011).

\bibitem{rodriguez}
I.D. Rodriguez, A. Sterdyniak, M. Hermanns, J.K. Slingerland, N. Regnault, Phys. Rev. B. {\bf 85} 035128 (2012)

\bibitem{yzz}
Bo Yang, Zi-Xiang Hu, Z. Papic and F. D. M. Haldane, Phys. Rev. Lett. {\bf 108}, 256807 (2012).

\bibitem{wen}
X.G. Wen and A. Zee, Phys. Rev. Lett. {\bf 69}, 953 (1992).

\bibitem{haldane}
F. D. M. Haldane, Phys. Rev. Lett. {\bf 51}, 605 (1983).

\bibitem{yzrh}
Bo Yang, Z. Papic, E.H. Rezayi, R.N. Bhatt and F.D.M. Haldane, Phys. Rev. B. {\bf 85}, 165318(2012).

\bibitem{bonderson}
Parsa Bonderson, Victor Gurarie and Chetan Nayak, Phys. Rev. B {\bf 83}, 075303 (2011)

\bibitem{yz}
Bo Yang, Z. Papic and F.D.M. Haldane, work in progress.

\bibitem{vignale}
I.V. Tokatly and G. Vignale, Phys. Rev. Lett. {\bf 98}, 026805 (2007).

\bibitem{jain2mode}
D. Majumder, S.S. Mandal and J.K. Jain, Nat. Phys. 5, 403 (2009).

\bibitem{wen}
X.G. Wen, Mod. Phys. Lett. {\bf B5}, 39 (1991).

\bibitem{stone}
M. Stone, H.W. Wyld and R.L. Schult, Phys. Rev. B. {\bf 45}, 14156 (1992).

\bibitem{readedge}
J. Dubail, N. Read and E.H. Rezayi, Phys. Rev. B. 86, 245310 (2012).

\bibitem{boyang}
Bo Yang, unpublished

\end{thebibliography}
\end{document}